\documentclass[aps,twocolumn]{revtex4}
\usepackage{epsfig}
\usepackage{amsfonts}

\begin{document}
\title
{Quantum Fokker-Planck-Kramers equation and entropy production}

\author{M\'ario J. de Oliveira}

\affiliation{
Instituto de F\'{\i}sica,
Universidade de S\~{a}o Paulo, \\
Rua do Mat\~ao, 1371,
05508-090 S\~{a}o Paulo, S\~{a}o Paulo, Brazil}

\date{\today}

\begin{abstract}

We use a canonical quantization procedure to set up 
a quantum Fokker-Planck-Kramers equation 
that accounts for quantum dissipation in a thermal
environment. The dissipation term is chosen to ensure that
the thermodynamic equilibrium is described by the Gibbs state.
An expression for the quantum entropy production is also
provided which properly describes quantum systems in 
a nonequilibrium stationary state. The time-dependent 
solution is given for a quantum harmonic oscillator in contact
with a heat bath. We also obtain the stationary solution for a
system of two coupled harmonic oscillators
in contact with reservoirs at distinct temperatures,
from which we obtain the entropy production and the
quantum thermal conductance.


\end{abstract}

\maketitle


Ordinary Brownian motion and other stochastic phenomena
with an underlying motion that follows the laws of classical mechanics  
are well described by the Fokker-Planck-Kramers (FPK) equation
\cite{kampen1981,gardiner1983,risken1984,tome2015book},
which accounts for the classical dissipative behavior in a thermal environment.
Quantum dissipation \cite{caldeira2014}, on the other hand,
cannot be explained by the classical FPK equation and
requires an extension of the stochastic theory to 
the domain of quantum mechanics. In fact, many
approaches to a stochastic theory of quantum systems have been put forward
\cite{caldeira2014,breuer2002,gardiner2010,weiss2012,
lindblad1976,iche1978,dekker1979a,dekker1979b,dekker1981,
caldeira1981,caldeira1983a,caldeira1983b,chang1985,gallis1993,
munro1996,banik2002}.
We mention the approach of
Lindblad \cite{lindblad1976}, which uses a
non-unitary time evolution of the density matrix, 
and the approach of Caldeira
and Leggett \cite{caldeira1983a,caldeira1983b} 
in which a quantum system interacts with a
heat reservoirs composed by a collection of harmonic
oscillators. 

The approach to a stochastic theory of quantum systems
that we consider here is based on a
quantum version of the classical FPK equation.
The construction of the quantum FPK equation
that we consider here is based on the canonical quantization
of the classical FPK equation. 
The resulting equation is similar to
the one found by Caldeira and Leggett \cite{caldeira1983b},
but differs by the dissipation term, which we
choose in such a way that in equilibrium 
the steady state is the Gibbs state. 

The classical FPK equation, 
when extended to a system of many particles, can also be the basis
for a stochastic approach to equilibrium and 
nonequilibrium thermodynamics
\cite{perez1994,mazur1999,tome2010,seifert2012,luposchainsky2013,tome2015}. 
Here, we also consider a quantum FPK equation
for a system with many degrees of freedom, which
is appropriate to describe a thermodynamic system
in contact with one or several thermal reservoirs,
which serves as the basis for a quantum thermodynamics 
\cite{bedeaux2001,kosloff2013}.
To this end it is necessary to define the rate of entropy production
for quantum system \cite{spohn1978,argentieri2014,carrillo2015}
which is also provided here.

The present approach is applied to two nonequilibrium situations.
In the first, we obtain the time-dependent
properties of a quantum harmonic oscillator in contact to a heat
reservoir. We find the time-dependent density matrix from which
we get the entropy production. In the second case, 
we apply the present approach to get the nonequilibrium steady state properties
of two coupled quantum oscillators in contact with two heat
reservoirs at distinct temperatures. We determine the 
entropy production as well as the heat flux across the
system.


We start with the classical FPK equation, which
gives the time evolution of the probability density
$P(x,p,t)$ related to the motion of a particle of mass $m$ subject to a 
potential $V(x)$ and in contact with a heat reservoir
at temperature $T$
\cite{kampen1981,gardiner1983,risken1984,tome2015book},
\begin{equation}
\frac{\partial P}{\partial t} = -\frac{p}{m}\frac{\partial P}{\partial x}
+\frac{dV}{dx}\frac{\partial P}{\partial p} 
+ \gamma \frac{\partial (pP)}{\partial p} + \frac{\gamma m}{\beta}
\frac{\partial^2 P}{\partial p^2},
\label{5}
\end{equation}
where $x$ and $p$ are the position and momentum of the particle,
$\gamma$ is the dissipation parameter and $\beta=1/k_BT$.
In the stationary state, it is straightforward to show that
$P=(1/Z)e^{-\beta{\cal H}}$, which means to say that indeed the
equation (\ref{5}) describes the contact with a heat bath.

Using the definition of the Poisson brackets
$\{A,B\} = (\partial A/\partial x)(\partial B/\partial p)
-(\partial A/\partial p)(\partial B/\partial x)$,
the FPK equation (\ref{5}) can be written in the form
\begin{equation}
\frac{\partial P}{\partial t} = \{{\cal H},P\}
+\gamma \{x,pP\} +\frac{\gamma m}{\beta} \{x,\{x,P\}\},
\label{7}
\end{equation}
where ${\cal H}=p^2/2m+V(x)$ is the Hamiltonian of the system.

A canonical quantization \cite{merzbacher1970} of equation (\ref{7})
can be achieved by replacing
the Poisson bracket $\{A,B\}$ by $[\hat{A},\hat{B}]/i\hbar$
where $[\hat{A},\hat{B}]=\hat{A}\hat{B}-\hat{B}\hat{A}$
is the commutator between the quantum operators $\hat{A}$
and $\hat{B}$ associated to the quantities $A$ and $B$, respectively.
For instance, the commutator between $x$ and $p$ is $[x,p]=i\hbar$.
This procedure combined with the use of a symmetrized product 
leads us to the following equation \cite{caldeira1983b} 
\begin{equation}
i\hbar\frac{\partial{\rho}}{\partial t} = [{{\cal H}},{\rho}]
+ \frac{\gamma}2 [{x},{\rho}{p}+{p}{\rho}]
+ \frac{\gamma m}{i\hbar\beta} [{x},[{x},{\rho}]],
\label{9}
\end{equation}
where ${\rho}$ is the density matrix and ${\cal H}={p}^2/2m+V({x})$
is the quantum Hamiltonian. 
This procedure will guarantee that, in the classical
limit, equation (\ref{9}) will become the FPK equation
(\ref{7}). However, as can be verified by a direct substitution,
$e^{-\beta{\cal H}}$ is not the stationary
solution of (\ref{9}), except for a free particle.
 
This simple procedure of replacing the classical Poisson
bracket by the quantum commutator and the classical variables by
quantum operators does not give an unambiguous
prescription to construct a quantum version of a classical
equation. Bearing this in mind, we look for a more general
quantum version of equation (\ref{7}) by assuming the 
following form for the quantum FPK equation 
\begin{equation}
i\hbar\frac{\partial{\rho}}{\partial t} = [{{\cal H}},{\rho}]
+ \frac{\gamma}2 [{x},{\rho}{g}+{g}^\dagger{\rho}]
+ \frac{\gamma m}{i\hbar\beta} [{x},[{x},{\rho}]],
\label{10}
\end{equation}
where the operator ${g}$ does not
depend on ${\rho}$ and are to be found
based on two assumptions. First, ${g}\to p$ 
in the classical limit,
so that equation (\ref{10}) goes onto (\ref{7}) in this limit.
Second, we require that ${\rho}_0=(1/Z) e^{-\beta{\cal H}}$
is the stationary solution of (\ref{10}) for any potential $V$.
To this end, we begin by writing equation (\ref{10}) in the form
\begin{equation}
i\hbar\frac{\partial{\rho}}{\partial t} = [{{\cal H}},{\rho}]
- [{x},{J}({\rho})],
\label{11}
\end{equation}
where
\begin{equation}
{J}({\rho}) = -\frac\gamma2({\rho}{g} + {g}^\dagger{\rho})
- \frac{\gamma\,m}{i\hbar\beta} [{x},{\rho}].
\label{12}
\end{equation}

In the stationary state, which is understood here as the thermodynamic
equilibrium state, ${J}=0$, that is, the insertion of the equilibrium
density matrix ${\rho}_0=(1/Z_0) e^{-\beta{\cal H}}$ into (\ref{12}) should
result in ${J}({\rho_0})=0$. This yields
\begin{equation}
{g}
= - \frac{ m}{i\hbar\beta}({\rho}_0^{-1}{x}{\rho}_0 - {x})
= - \frac{ m}{i\hbar\beta}(e^{\beta{\cal H}}{x}e^{-\beta{\cal H}} - {x}),
\label{14}
\end{equation}
which is the desired expression for $g$.
The expansion of the first term between parentheses in powers of
$\beta$
allows us to write $g$
in a form involving nested commutators
\begin{equation}
{g} = {p} + \frac{\beta}{2!}[{\cal H},{p}]
+ \frac{\beta^2}{3!}[{\cal H},[{\cal H},{p}]]
+ \frac{\beta^3}{4!}[{\cal H},[{\cal H},[{\cal H},{p}]]] + \ldots
\label{15}
\end{equation}
In the classical limit, all terms, except the first, on
the right-hand side of equations (\ref{15})
vanish and ${g}$ approaches $p$ as desired.


Next we wish to connect the present approach
with nonequilibrium thermodynamics. To this end we
consider the evolution of the free energy $F$, defined
by $F=U-TS$, where $U=\langle{\cal H}\rangle$ and $S$ is the
von Neumann entropy, $S=-k_B{\rm Tr}\{\rho\ln\rho\}$.
The expression for the free energy can be written in the form
\begin{equation}
F = k_B T\, {\rm Tr}\{\rho\ln\rho -\rho\ln\rho_0 \} + F_0,
\label{17}
\end{equation}
where $\rho_0=(1/Z)e^{-\beta{\cal H}}$ is the equilibrium density matrix
and $F_0=-k_BT\ln Z_0$. Using Klein's inequality \cite{klein1931},
${\rm Tr}\{\rho\ln\rho -\rho\ln\rho_0 \}\geq0$, 
it follows at once that $F\geq F_0$.

The time derivative of $F$ is related to the
entropy production rate $\Pi$ by $dF/dt=-T\Pi$ \cite{tome2010,tome2015}. 
Using equation (\ref{11}) to calculate 
$dF/dt$ from equation (\ref{17}), we arrive at the following expression
for the rate of entropy production,
\begin{equation}
\Pi = \frac{k_B}{i\hbar}{\rm Tr}\{[x,J(\rho)](\ln \rho - \ln \rho_0)\},
\label{19a}
\end{equation}
which can also be written as
\begin{equation}
\Pi = \frac{k_B}{i\hbar}{\rm Tr}\{[x,J(\rho)](\ln \rho +\beta{\cal H}\}.
\label{19}
\end{equation}

It is worth mentioning, that, in the classical limit,
the entropy production rate reduces to 
the following expression \cite{tome2010,tome2015}
\begin{equation}
\Pi = \frac{1}{\gamma\, Tm}\int \frac{J^2}{P}dxdp.
\end{equation}
which is a quantity manifestly nonnegative, where
$J=-\gamma pP-(\gamma m/\beta)\partial P/\partial p$.


Let us determine ${g}$ for some simple situations.
In the case of a free particle, for which ${\cal H}={p}^2/2m$,
the quantity ${g}$ simplifies substantially. In this
case all the commutators in equations (\ref{15}) vanish and
${g}={p}$. The resulting quantum FPK equation
reduces to equation (\ref{9}), which is thus understood as
the equation describing a quantum Brownian motion of a free particle.


In the case of an harmonic oscillator, for which
${\cal H}={p}^2/2m + m\omega^2{x}^2/2$,
a straightforward calculation gives $g = a p + i b x$,
where $a$ and $b$ are real numbers, given by
\begin{equation}
a = \frac{1}{\beta\hbar\omega}\sinh\beta\hbar\omega,
\qquad
b = \frac{m}{\beta\hbar}(\cosh\beta\hbar\omega - 1).
\label{16}
\end{equation}
The time evolution of the covariances are
\begin{equation}
\frac{d}{dt}\langle p^2\rangle = 
-m\omega^2 (\langle px\rangle + \langle xp\rangle) + \hbar b\gamma
-2a\gamma \langle p^2\rangle + \frac{2\gamma m}{\beta},
\label{23a}
\end{equation}
\begin{equation}
\frac{d}{dt}\langle x^2\rangle = \frac{1}{m} (\langle px\rangle + \langle xp\rangle),
\label{23b}
\end{equation}
\begin{equation}
\frac{d}{dt}\langle px\rangle =
\frac{d}{dt}\langle xp\rangle = \frac{1}{m} \langle p^2\rangle
-m\omega^2\langle x^2\rangle
-\frac{a\gamma}2(\langle px\rangle + \langle xp\rangle).
\label{23c}
\end{equation}
At the stationary state, 
$\langle xp\rangle =-\langle xp\rangle=i\hbar/2$,
$\langle p^2\rangle=(\hbar b/2a)+(m/a\beta)$, and
$\langle x^2\rangle=\langle p^2\rangle/m\omega^2$.
From these results one gets the expected expression for $\langle{\cal H}\rangle$,
\begin{equation}
\langle{\cal H}\rangle = \hbar\omega\left(
\frac{1}{e^{\beta\hbar\omega}-1}+\frac12\right).
\end{equation}

\begin{figure}
\epsfig{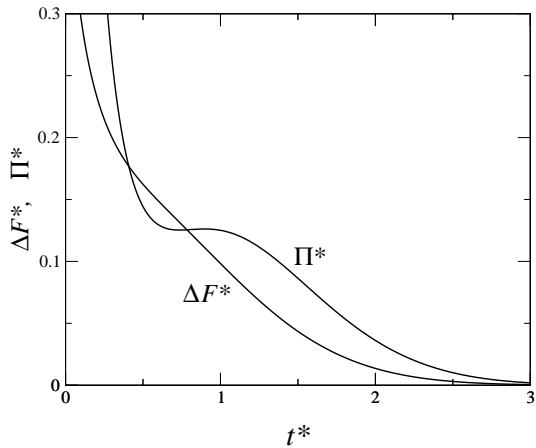}
\caption{Free energy and rate of entropy production as functions of time
for a quantum oscillator in contact with a heat reservoir.
The dimensionless quantities in the plot are as follows:
$\Delta F^*=(F-F_0)/\hbar\omega$, $\Pi^*=\Pi/k_B\omega$ and
$t^*=\omega t$. The values of the parameters are: $k_B T/\hbar\omega=1$,
$\gamma/\omega=1$. }
\label{fent}
\end{figure}

A time dependent solution of the quantum FPK equation for the harmonic
oscillator is
\begin{equation}
\rho = \frac1Z \exp\{-\frac{c_1}2 p^2 - \frac{c_2}2 x^2 - \frac{c_3}2(xp+px)\},
\label{35}
\end{equation}
where $c_1$, $c_2$ and $c_3$ are time-dependent parameters.
That this form is indeed a solution can be verified by replacing
(\ref{35}) into the FPK equation. From (\ref{35}) one obtains
the relation between the covariances $\langle p^2\rangle$, $\langle x^2\rangle$,
$\langle xp+px\rangle$ and the parameters $c_1$, $c_2$, $c_3$ so that
from the time-dependent solution of equations (\ref{23a}), (\ref{23b}),
and (\ref{23c}), we may find the time behavior of
$c_1$, $c_2$, $c_3$, and $\rho$.
We are thus able to get the time-dependent properties of the
harmonic oscillator in contact with a heat reservoir, given an initial
condition. Using this procedure, we have determined the time behavior
of several quantities, including
the free energy $F$ and the production of entropy $\Pi$, which are
shown in figure \ref{fent}. In this figure we used initial conditions
such that $\langle p^2\rangle/m\hbar\omega=1/2$,
$m\langle x^2\rangle/\hbar\omega=1/2$, and $\langle xp+px\rangle=0$.

As stated before, the present approach may serve as the basis
for a nonequilibrium thermodynamics. The fundamental properties
that distinguish nonequilibrium from equilibrium are the 
production of entropy, whose nonnegativity is equivalent to the
second law, and the existence of fluxes such as heat flow. 
It is convenient to distinguish two forms of nonequilibrium.
A system may be out of equilibrium because it did not yet reach equilibrium.
This was exemplified by the case of an harmonic oscillator
that we have just considered.
As shown in figure \ref{fent}, the entropy production is nonnegative
but vanishes for large times because the system reaches equilibrium.
Another aspect is the nonequilibrium behavior in systems that are in 
the steady state, as happens to a system in contact to 
two heat reservoirs at distinct temperatures. In this case there
will be a permanent heat flow and the production of entropy is 
positive and constant in time. To deal with this case we consider
next a system of many interacting particles in contact with 
heat reservoirs at distinc temperatures.

A system of interacting particles is described by the Hamiltonian
\begin{equation}
{\cal H} = \sum_i \frac{p_i^2}{2m} + V(x),
\end{equation}
where here $x$ represents the collection of the positions $\{x_i\}$
of the particles and $p_i$ is the momentum conjugate to $x_i$.
The quantum FPK equation for this case reads
\begin{equation}
i\hbar\frac{\partial{\rho}}{\partial t} = [{{\cal H}},{\rho}]
-  \sum_i [{x}_i,J_i(\rho)],
\label{20}
\end{equation}
where $\rho(x,v)$ is the density matrix and
\begin{equation}
J_i(\rho) = -\frac{\gamma_i}2({\rho}{g}_i + {g}_i^\dagger{\rho})
- \frac{\gamma_i m}{i\hbar\beta_i}[{x_i},{\rho}],
\label{21}
\end{equation}
with
\begin{equation}
{g}_i = - \frac{ m}{i\hbar\beta_i}
(e^{\beta_i{\cal H}}{x}_ie^{-\beta_i{\cal H}} - {x}_i).
\end{equation}
We are considering a general case in which each particle
is in contact with a heat reservoir at a temperature $T_i$
and $\beta_i=1/k_BT_i$.

Let us assume that some particles are in contact ($\gamma_i\neq0$) with
the same reservoir at temperature $T$ whereas the others are
not connected ($\gamma_i=0$) to any reservoir. In this case,
the steady state solution of the quantum FPK equation (\ref{20})
is the canonical Gibbs density matrix
$\rho_0=(1/Z)e^{-\beta{\cal H}}$ because $J_i(\rho_0)=0$
for each $i$, a condition that may be understood as detailed balance,
and the system is found to be in thermodynamic equilibrium.
It is worth mentioning that the canonical density matrix $\rho_0$
will be the steady state solution no matter how many particles
are in contact ($\gamma_i\neq0$) with the reservoirs,
as long as there is at least one. This feature distinguishes
the present approach from that in which local Lindblad forms
are used to describe the contact with heat reservoirs, which
does not lead to a thermalization of the system into the Gibbs
state \cite{asadian2013}. 

If the temperatures of the heat reservoirs are different from each other
then, in the steady state, the system will not be in 
thermodynamic equilibrium. This nonequilibrium steady state
may be characterized by a nonzero production of entropy.
In analogy with equation (\ref{19}),
we define the rate of entropy production as
\begin{equation}
\Pi = \frac{k_B}{i\hbar}\sum_i {\rm Tr}\{[x_i,J_i(\rho)]
(\ln \rho + \beta_i{\cal H})\}.
\end{equation}
The time variation of the von Neumann entropy $S=-k_B{\rm Tr} \{\rho\ln \rho\}$
can be written as
\begin{equation}
\frac{dS}{dt} = \frac{k_B}{i\hbar} \sum_i {\rm Tr} \{[x_i,J_i(\rho)]\ln \rho\},
\end{equation}
so that the flux of entropy, $\Phi=\Pi-dS/dt$, from the system
toward the reservoirs is thus
\begin{equation}
\Phi = \frac{1}{i\hbar}\sum_i\frac{1}{T_i} {\rm Tr}\{[x_i,J_i(\rho)]{\cal H}\}.
\label{29}
\end{equation}

From the quantum FPK (\ref{20}), the evolution of the average
of the energy $U=\langle{\cal H}\rangle$ is given by
\begin{equation}
\frac{dU}{dt} = - \sum_i \phi_i,
\end{equation}
where 
\begin{equation}
\phi_i = \frac{1}{i\hbar} {\rm Tr}\{[x_i,J_i] {\cal H}\}
\end{equation}
is the heat flux from the system toward the reservoir at temperature $T_i$,
so that we may write 
\begin{equation}
\Phi = \sum_i \frac{\phi_i}{T_i}.
\end{equation}
Replacing $J_i$ in the expression for $\phi_i$, we get
\begin{equation}
\phi_i = \gamma_i
\left( \frac1{2m}\langle g_i p_i + p_i g_i^\dagger\rangle - k_B T_i\right).
\label{31}
\end{equation}
For two reservoirs in the 
stationary state, $\phi_1+\phi_2=0$ and
\begin{equation}
\Phi = \frac{\phi_1}{T_1}+\frac{\phi_2}{T_2}
=\phi\left(\frac{1}{T_2}-\frac{1}{T_1}\right),
\label{27}
\end{equation}
where $\phi=\phi_2=-\phi_1$ is interpreted as the heat flow
across the system from reservoir 1 to reservoir 2.


Let us consider a system of two coupled harmonic oscillators,
described by the Hamiltonian
\begin{equation}
{\cal H} = \frac{1}{2m}(p_1^2 + p_2^2) + \frac{k}2(x_1-x_2)^2 
+ \frac{k'}{2}(x_1^2 + x_2^2),
\end{equation}
in contact with reservoirs at temperature $T_1$ and $T_2$.
In this case the quantities $g_1$ and $g_2$ are given by
\begin{equation}
g_i = \sum_{j=1,2} (a_{ij}p_j + ib_{ij}x_j), 
\label{33}
\end{equation}
where 
\begin{equation}
a_{ij} = \frac{\sinh(\beta_i\hbar\omega)}{2\beta_i\hbar\omega}
\pm \frac{\sinh(\beta_i\hbar\omega')}{2\beta_i\hbar\omega'},
\end{equation}
\begin{equation}
b_{ij} = \frac{m\cosh(\beta_i\hbar\omega)-1}{2\beta_i\hbar}
\pm \frac{m\cosh(\beta_i\hbar\omega')-1}{2\beta_i\hbar},
\end{equation}
where the plus and minus signs are to be used when $i=j$ and $i\neq j$,
respectively, and $\omega=\sqrt{k'/m}$ and $\omega'=\sqrt{(2k+k')/m}$.

From the quantum FPK equation (\ref{20}), and using $g_i$ given by
(\ref{33}), we set up the evolution equations for the correlations
$\langle x_ix_j\rangle$, $\langle x_ip_j\rangle$, and $\langle p_ip_j\rangle$.
From the solution of these equations we can determine the
heat flow $\phi$, given by (\ref{31}). In the stationary state, we find
\begin{equation}
\phi = \frac{k^2}{k'm\gamma(a_1 + a_2)}\,
\left(\frac{\hbar\omega}{e^{\beta_1\hbar\omega}-1}
- \frac{\hbar\omega}{e^{\beta_2\hbar\omega}-1}\right),
\label{36}
\end{equation}
valid for $k<<k'$, where
\begin{equation}
a_1 = \frac{\sinh(\beta_1\hbar\omega)}{\beta_1\hbar\omega},
\qquad\qquad
a_2 = \frac{\sinh(\beta_2\hbar\omega)}{\beta_2\hbar\omega},
\end{equation}
We have used the same dissipation parameters, $\gamma_1=\gamma_2=\gamma$.
The heat flux $\phi$, given by equation (\ref{36}), is positive if $T_2>T_1$ and
negative if $T_2<T_1$. Therefore, the entropy production $\Pi$, which
in the stationary state is identified as the entropy flux $\Phi$,
given by (\ref{27}), is nonnegative, as desired.
Notice that, in the classical limit, expression (\ref{36}) 
reduces to the expression that one obtains
from the classical FPK equation \cite{tome2010,morgado2009}.

The quantum thermal conductance $\kappa$ is obtained by writing
$T_{1,2}=T\pm \Delta T/2$ and $\phi=\kappa \Delta T$.
For small values of $\Delta T$,
\begin{equation}
\kappa = \frac{k_B k^2}{2k'm\gamma a}\,
\left(\frac{\beta\hbar\omega}{e^{\beta\hbar\omega}-1}\right)^2.
\label{37}
\end{equation}
where $a$ is given by (\ref{16}).


In summary, by the use of a canonical quantization we have set up 
a quantum FPK equation describing the time evolution
of quantum systems in contact with heat reservoirs.
For a system in contact with just one heat reservoir
the stationary state is the equilibrium Gibbs state.
We have applied the present approach to a system
of two coupled harmonic oscillators in contact with
reservoirs at distinct temperatures. From the steady
state solution of the quantum FPK equation we have
obtained the heat flux and the entropy production
which was shown to be positive.

As a final comment, it should be pointed out that the
results we have obtained for the harmonic oscillator
in equilibrium do not depend on damping parameter.
Since this is to be expected when the damping parameter
is small \cite{caldeira2014}, it follows that
the present approach 
is able to give the expected results in the quantum regime at
least in the low damping regime.

We wish to acknowledge useful conversation with G. T. Landi.



\begin{thebibliography}{99}

\bibitem{kampen1981} N. G. van Kampen, {\it Stochastic Processes in Physics
and Chemistry}, North-Holland, Amsterdam, 1981.

\bibitem{gardiner1983} C. W. Gardiner, {\it Handbook of
Stochastic Methods for Physics, Chemistry and Natural Sciences},
Springer, Berlin, 1983.

\bibitem{risken1984} The Fokker-Planck Equation, {\it Methods of 
Solution and Applications}, Springer, Berlin, 1984.

\bibitem{tome2015book} T. Tom\'e and M. J. de Oliveira,
{\it Stochastic Dynamics and Irreversibility}, Springer,
Heidelberg, 2015.


\bibitem{caldeira2014} A. O. Caldeira, {\it 
An Introduction to Macroscopic Quantum Phenomena and Quantum Dissipation},
Cambridge University Press, 2014.

\bibitem{breuer2002} H.-P. Breuer and F. Petruccione,
{\it The Theory of Open Quantum System},
Oxford University Press, 2002.

\bibitem{gardiner2010} C. W. Gardiner and P. Zoller,
{\it Quantum Noise}, Springer, 2010. 3rd ed. 

\bibitem{weiss2012} U. Weiss, {\it Quantum Dissipative Systems},
World Scientific, 2012, 4th ed.


\bibitem{lindblad1976} G. Lindblad,
Commun. Math. Phys. {\bf 48}, 119 (1976).

\bibitem{iche1978} G. Iche and P. Nozi\`eres, Physica A {\rm 91}, 485 (1978).

\bibitem{dekker1979a} H. Dekker, Physica A {\bf 95}, 311 (1979).

\bibitem{dekker1979b} H. Dekker, Phys. Lett. A {\bf 74}, 15 (1979).

\bibitem{dekker1981} H. Dekker, Physics Report {\bf 80}, 1 (1981).

\bibitem{caldeira1981} A. Caldeira and A. J. Leggett,
Phys. Rev. Lett. {\bf 46}, 211 (1981).

\bibitem{caldeira1983a} A. O. Caldeira and A. J. Leggett, 
Ann. Phys. {\bf 149}, 374 (1983).

\bibitem{caldeira1983b} A. O. Caldeira and A. J. Leggett, 
Physica A {\bf 121}, 587 (1983).

\bibitem{chang1985} L.-D. Chang and D. Waxman, 
J. Phys. C {\bf 18}, 5973 (1985).

\bibitem{gallis1993} M. R. Gallis, Phys. Rev. A {\bf 48}, 1028 (1993).

\bibitem{munro1996} W. J. Munro and C. W. Gardiner,
Phys. Rev. A {\bf 53}, 2633 (1996).

\bibitem{banik2002} S. K. Banik, B. C. Bag, and D. S. Ray,
Phys. Rev. E {\bf 65}, 051106 (2002).


\bibitem{perez1994} A. P\'erez-Madrid, J. R. Rub\'{\i} and P. Mazur,
Physica A {\bf 212}, 231 (1994).

\bibitem{mazur1999} P. Mazur, Physica A {\bf 274}, 491 (1999).

\bibitem{tome2010} T. Tom\'e and M. J. de Oliveira,
Phys. Rev. E {\bf 82}, 021120 (2010).

\bibitem{seifert2012} U. Seifert, Rep. Prog. Phys. {\bf 75},
126001 (2012).

\bibitem{luposchainsky2013} D. Luposchainsky and H. Hinrichsen,
J. Stat. Phys. {\bf 153}, 828 (2013).

\bibitem{tome2015} T. Tom\'e and M. J. de Oliveira,
Phys. Rev. E {\bf 91}, 042140 (2015).


\bibitem{bedeaux2001} D. Bedeaux and P. Mazur, Physica A {\bf 298}, 81 (2001).

\bibitem{kosloff2013} R. Kosloff, 
Entropy {\bf 15}, 2100 (2013).

\bibitem{spohn1978} H. Spohn, J. Math. Phys. {\bf 19}, 1227 (1978).

\bibitem{argentieri2014} G. Argentieri, F. Benatti, R. Floreanini, and M. Pezzutto,
arXiv:1408.4589


\bibitem{carrillo2015} E. Solano-Carrillo and A. J. Mills,
arXiv:1509.04635


\bibitem{merzbacher1970} E. Merzbacher, {\it Quantum Mechanics},
Wiley, New York, 1970, 2nd ed.

\bibitem{klein1931} O. Klein, 
Z. Physik {\rm 72}, 767 (1931).

\bibitem{asadian2013} A. Asadian, D. Manzano, M. Tiersch, and H. J. Briegel,
Phys. Rev. E {\bf 87}, 012107 (2013).

\bibitem{morgado2009} W. A. M. Morgado and D. O. Soares-Pinto,
Phys. Rev. E {\bf 79}, 051116 (2009). 

\end{thebibliography}
\end{document}